\documentstyle[aps]{revtex}
\begin{document}
\draft
\preprint{hep-ph/9504???}

\title{Expansion, Thermalization and Entropy Production\\
in High-Energy Nuclear Collisions}
\author{H. Heiselberg}
\address{NORDITA, Blegdamsvej 17, DK-2100 Copenhagen \O., Denmark}
\author{ Xin-Nian Wang}
\address{ Nuclear Science Division, MS 70A-3307,
         Lawrence Berkeley Laboratory, \\
         University of California, Berkeley, California 94720, USA }
\maketitle

\begin{abstract}
\baselineskip=15pt
The thermalization process is studied in an expanding parton gas  using the
Boltzmann equation with two types of collision terms. In the relaxation time
approximation we determine the criteria  under which a time-dependent
relaxation time leads to thermalization of the partons. We calculate the
entropy production due to collisions for the general time-dependent relaxation
time. In a perturbative QCD approach on the other hand, we  can estimate  the
parton collision time and its dependence on  expansion time. The effective
`out of equilibrium' collision time  differs from the standard transport
relaxation time,  $\tau_{\rm tr}\simeq(\alpha_s^2\ln(1/\alpha_s)T)^{-1}$, by a
weak time  dependence. It is in both cases
Debye screening and Landau damping that
regulate the singular forward scattering processes. We find that the parton
gas does thermalize eventually but only after having undergone a phase of free
streaming and gradual equilibration where considerable entropy is produced
(``after-burning"). The final entropy and thus particle density depends on the
collision time as well as the initial conditions (a ``memory effect"). Results
for entropy production are presented based upon various model estimates of
early parton production.
\end{abstract}

\baselineskip=16pt
\section{Introduction}

Hard and semihard parton scatterings are expected to be the dominant processes
in ultrarelativistic heavy ion collisions at the Relativistic Heavy Ion
Collider (RHIC) and Large Hadron Collider (LHC) energies \cite{JBAM,KLL}.
These hard or semihard processes happen in a very short time scale after which
a dense parton gas will be formed. However, this parton gas is not immediately
in thermal and chemical equilibrium. Secondary interactions among the produced
partons may lead to equilibrium if the interactions are sufficiently strong.
Exactly how this parton gas equilibrate and what  are the time scales are
under intense investigations  \cite{Geiger,Biro,Wang,LXES} and that is also the
focus of this paper.

It is commonly assumed in relativistic heavy ion collisions that the matter
expands hydrodynamically shortly after the nuclei collide and that it is in
thermal equilibrium locally in space and time. In the classic Bjorken model
\cite{Bjorken} hydrodynamic expansion is assumed after a time $\tau_0\simeq
1$fm/c. Yet, when initial parton formation times are estimated to be a
fraction of a fm/$c$ \cite{Geiger,Wang,Shuryak} the expansion might be much
more rapid than the typical collision time of the partons produced in the
collisions. For times shorter than the typical collision time of partons
immediately after the initial collisions, the rapid expansion is closer to
free streaming than hydrodynamic expansion. Only at times much larger than
the characteristic collision time may the parton gas thermalize, equilibrate
and expand hydrodynamically \cite{Baym}. However, if the collision time
increases with time the gas may never thermalize. The characteristic collision
time is therefore a crucial parameter. Since it among other things depends on
the density, which decreases in time, it may be time dependent. In fact, when
the collision time is proportional to the expansion time, the system neither
expands hydrodynamically nor as free streaming at large times but somewhere
in between \cite{Gavin}.

In this paper we study the thermalization process of a parton gas expanding
in one dimension using the Boltzmann equation with two types of collision
terms. In the relaxation time approximation we show generally for which
time-dependent relaxation times the partons thermalize and estimate when they
do so and how much entropy is produced in the collisions. Secondly, we
describe the parton scatterings by perturbative QCD in order to estimate the
important collision time and its dependence on expansion time. The effective
``out of equilibrium'' collision time differ from the standard transport
relaxation time, $\tau_{\rm tr}\simeq(\alpha_s^2\ln(1/\alpha_s)T)^{-1}$, by a
prefactor and a weak time dependence, which aer calculated in the Appendix,
It is, however, still the Debye screening and Landau damping that screen the
singular forward parton scattering processes. Finally, we will give results
for entropy production based upon various model estimates of the
preequilibrium condition.

\section{Free Streaming vs. Hydrodynamic Expansion}

In order to have a tractable approach we assume that the spatial variations
are sufficiently small along the longitudinal (z) direction so that we can
describe the parton gas by the Boltzmann equation
\begin{eqnarray}
   \left(\frac{\partial}{\partial t}+{\bf v}_{\bf p}\cdot\nabla_{\bf r}
   \right) f  = \left(\frac{\partial f}{\partial t}\right)_{\rm coll}
     \, . \label{BE}
\end{eqnarray}

We also assume along the lines of the Bjorken model \cite{Bjorken} that the
transverse dimension of the parton gas formed in the nuclear collisions is
sufficiently large that the initial expansion is one-dimensional. Furthermore,
we assume  the central rapidity regime is boost invariant in rapidity,  {\it
i.e.}, depending only on the invariant time  $\tau=\sqrt{t^2-z^2}$. The
Boltzmann equation thus  reduces to \cite{Baym}
\begin{eqnarray}
\left(\frac{\partial f(p_\perp,p_z,\tau)}{\partial\tau}\right)_{|_{p_z\tau}} =
\left(\frac{\partial f(p_\perp,p_z,\tau)}{\partial\tau} \right)_{\rm coll}
\, . \label{fscale}
\end{eqnarray}

The collision term on the r.h.s. of Eqs.~(\ref{BE},\ref{fscale}) determines
the equilibration and has been studied extensively within perturbative
QCD \cite{DG,HK,BP1,BP2,eta,THOMA} near equilibrium in a
quark-gluon plasma. In a relaxation time approximation, the collision
term is
\begin{eqnarray}
 \left(\frac{\partial f(p_\perp,p_z,\tau)}{\partial\tau} \right)_{|_{p_z\tau}}
   = \, - \, \, \frac{f-f_{\rm eq}}{\theta}    \, , \label{relax}
\end{eqnarray}
where $\theta$ is the relaxation time, which determines the time scale
for thermalization and $f_{\rm eq}$ is the equilibrium distribution function.

\subsection{The hydrodynamic limit}

When collisions are sufficient to thermalize the system, {\it i.e.},
the corresponding relaxation times are short as compared to
expansion times, $\theta\ll\tau$, hydrodynamics applies and the
distribution function in the local comoving frame is the thermal one
\begin{eqnarray}
    f_{\rm eq}({\bf p}) = (\exp(\frac{E_p-\mu}{T})\pm 1)^{-1}
    \, .\label{fthermal}
\end{eqnarray}
If particle production is sufficiently rapid, chemical equilibrium
can also be reached and $\mu\simeq0$ for gluons. We assume that the net baryon
density is relatively low in the midrapidity region so that the
quark chemical potential vanishes as well. In the Bjorken flow
model \cite{Bjorken} the expansion is idealized in one dimension
and assumed to be rapidity independent. Including viscous dissipation
one finds generally in hydrodynamics \cite{DG,HK,Gavin2}
\begin{eqnarray}
   \frac{d\epsilon}{d\tau} + \frac{\epsilon+P}{\tau} =
         \frac{\frac{4}{3}\eta+\xi}{\tau^2} \, , \label{Hydro}
\end{eqnarray}

where $\epsilon$ is the energy density, $P$ the pressure, $\eta$ and $\xi$ the
shear and bulk viscosities respectively. We assume that parton gas is a
weakly interacting QGP consisting of relativistic quarks and gluons.
Consequently, the gas is described by an ideal equation of state  with
pressure $P=\epsilon/3$ where the energy density is $\epsilon=aT^4$;
$a=\frac{8\pi^2}{15}(1+\frac{21}{32}N_f)$ ($a=15.6$ for $N_f=f$). A bag
pressure can be included but will not change the entropy production that we
are concerned with here.

In the relaxation time approximation the shear viscosity is $\eta=\theta
\epsilon$. In a weakly interacting  QGP the temperature is the only scale
besides factors of the interaction strengths, {\it e.g.}, the Debye wavenumber
is $q_D\sim gT$. Thus the shear viscosity necessarily scales like
\begin{eqnarray}
   \eta =\tilde{\eta}T^3 \, ,
\end{eqnarray}
where $\tilde{\eta}$ is a dimensionless constant depending on the coupling
constant $g$. The bulk viscosity $\xi$ vanishes for an ideal  relativistic gas
\cite{xi}.

Solving Eq.~(\ref{Hydro}), the entropy density $s=(4/3)\epsilon/T$ is obtained.
The total entropy, $S$, is proportional to $\tau s$  and is thus \cite{DG,HK}
\begin{eqnarray}
   S = S_0 \left( 1+\frac{\tilde{\eta}}{2a\tau_0T_0}
       (1 - (\frac{\tau_0}{\tau})^{2/3})
      \right)^3 \, , \label{S}
\end{eqnarray}
where $S_0$, $\tau_0$ and $T_0$ are the initial entropy, time and temperature.
As seen in Fig.~\ref{fig1} the entropy production increases rapidly
with viscosity $\tilde{\eta}$. If the decoupling (freeze-out)
or hadronization time is sufficiently late the entropy approaches the
asymptotic value $S=S_0(1+\tilde{\eta}/2a\tau_0T_0)^3$.

To get an idea of the magnitude of $\tilde{\eta}$ we refer
to recent calculation within perturbative QCD \cite{BP1,eta}. For weak
couplings, $\alpha_s\raisebox{-.5ex}{$\stackrel{<}{\sim}$} 0.1$, the
quark and gluon viscosities depend on the coupling constant as
\begin{eqnarray}
   \tilde{\eta}\simeq (\alpha_s^2\ln(1/\alpha_s))^{-1} \, ,
\end{eqnarray}
where the logarithm arises from
the sensitivity to screening of long range (small momentum transfer)
interactions. The quark viscosity is approximately four times larger than the
gluon because quarks interact more weakly. For strongly interacting plasmas,
$\alpha_s\raisebox{-.5ex}{$\stackrel{>}{\sim}$} 0.1$, the Debye and dynamical
screening must be replaced by some
effective cutoff due to correlations in the plasma, which leads to
$\tilde{\eta}\propto \alpha_s^{-2}$.
Within the range $\alpha_s\sim 0.1-0.5$, the quark (gluon) viscosity
decrease from $\tilde{\eta}\sim 50$ (15) to $\tilde{\eta}\sim 1.5$ (0.8).
If the initial density of quarks and gluons, $n_i$, is less than that in
chemical equilibrium, $n_{eq}$, then the viscosities are larger by a factor
$\sim n_{eq}/n_i$.

The corresponding viscous relaxation time
\begin{eqnarray}
 \tau_\eta=5\frac{\eta}{\varepsilon+P}=\frac{15}{4a}\tilde{\eta}T^{-1}  \, ,
\end{eqnarray}
is, however, larger than the expansion time and so hydrodynamics does not
apply at early times as pointed out by Danielewicz and Gyulassy \cite{DG}. The
entropy production will be much lower and will be calculated in the following
sections by solving the Boltzmann equation in the relaxation time
approximation. For comparison the upper limit on the entropy production in
Eq.(\ref{sbaym}) from solving the Boltzmann equation with a  finite collision
time, $\theta=\tau_\eta$, for the case $\tilde{\eta}=16$ corresponding to
$\theta\simeq 4\tau_0$ is also shown in Fig.~\ref{fig1} as a dashed line. Note
that the  entropy production continues long after $\theta$ and $\tau_\eta$.

\subsection{Free Streaming}
In the opposite extreme to the hydrodynamic limit examined above,
collisions are absent and partons stream freely. According
to Eq.~(\ref{fscale}) the distribution function evolves
as \cite{Baym}
\begin{eqnarray}
     f({\bf p}) = f_0({\bf p}_\perp,p_z\tau/\tau_0) \, ,\label{fstream}
\end{eqnarray}
where $f_0$ is the initial distribution function at time $\tau_0$.

As will be described in the following section  the parton gas will stream
freely until collisions thermalize the system around a time $\theta$.  The
free streaming alters the distribution in phase space drastically. Initially
the partons have large longitudinal momenta due to the high relative energy of
the incoming nucleons in the nucleus-nucleus collisions and the relatively
small transverse momentum transfer in hadronic collisions. However, in the
one-dimensional free streaming expansion of the system at later times only
those partons with similar longitudinal velocity will  travel together locally
in space and time. Thus the phase space separates the longitudinal momenta and
the distribution function changes from a wide to a narrow one in $p_z$ locally
in space and time (see Fig. \ref{fig4}). Collisions will then attempt to
thermalize the system towards an isotropic distribution.

When the longitudinal expansion has extended the system to a size similar to
the transverse size, which happens at a time of order the nuclear transverse
dimension $\tau\sim R$, three dimensional expansion takes over. Hereafter the
densities will decrease rapidly reducing collisions drastically and a free
streaming scenario is again likely. Around the same time, however, the system
may break up, freeze-out and fragment.

\section{Thermalizaton in the Relaxation Time Approximation}

Baym solved the Boltzmann equation in the relaxation time approximation
Eq.~(\ref{relax}) with a relaxation time $\theta$ {\it independent} of time
\cite{Baym}.   He found that the parton gas started out free streaming and
gradually thermalized to hydrodynamical flow on a time scale  given by
$\theta$. The distribution function at any point in the  local rest frame
changes from being highly anisotropic, with only  small longitudinal momenta,
to being isotropic for $\tau\gg\theta$.  A similar calculation by Gavin
\cite{Gavin}, however, with a collision time scaling {\it linearly} with time,
$\theta=\alpha t$, gave a  qualitatively different result. In this case the
gas ends up somewhere between free streaming and thermal equilibrium depending
on the values of $\alpha$. For small $\alpha$ the collision  time is short and
the  final state is close to Bjorken flow but for large $\alpha$ the state  is
closer to free streaming. The time dependence and magnitude  of the relaxation
time is thus essential for describing the degree  of thermalization and its
time scale.

We will in the following solve the Boltzmann equation within the
relaxation time approximation for a more general time dependence
of the collision time proportional to the expansion time $\tau$
to some power,
\begin{eqnarray}
   \theta \propto \tau^p \, . \label{power}
\end{eqnarray}
This covers the constant relaxation time of Baym \cite{Baym} ($p=0$),
the linear one of Gavin \cite{Gavin} $\tau_{coll}=1/n\sigma\sim\tau$
(i.e. $p=1$) as well as
the near equilibrium transport relaxation time
$\tau_{\rm tr}\sim T^{-1}\sim n^{-1/3}\sim\tau^{1/3}$ (i.e. $p=1/3$).
We shall study under which circumstances the parton gas thermalizes,
estimate the relaxation times and predict
the degree and time scale for equilibration.

The solution to the Boltzmann equation (\ref{relax}) can be
written in terms of an integral equation
\begin{eqnarray}
    f&=&f_0(p_\perp,p_z\tau/\tau_0)e^{-x} + \int^{x}_0 dx'
    e^{x'-x}f_{\rm eq}(\sqrt{p_\perp^2+(p_z\tau/\tau')^2},T',\mu')
     \, , \label{fint}
\end{eqnarray}
where the time dependence of the temperature $T$ and chemical
potential $\mu$ in $f_{\rm eq}(p,T,\mu)$ are determined by
demanding the energy density and number density (assuming
no particle production) be the same for $f_{\rm eq}$ and $f$
at any time, {\it i.e.},
\begin{eqnarray}
   \epsilon(T,\mu) &=& \int d\Gamma_{\bf p} E_p f({\bf p})
               \equiv \int d\Gamma_{\bf p} E_p f_{\rm eq}(p,T,\mu)
               \, , \label{cond1} \\
   n(T,\mu) &=& \int d\Gamma_{\bf p} f({\bf p})
               \equiv \int d\Gamma_{\bf p} f_{\rm eq}(p,T,\mu) \,
               , \label{cond2}
\end{eqnarray}
where $d\Gamma_{\bf p}=d^3{\bf p}/(2\pi)^3$.
The function $x(\tau)$ is related to $\tau$ by
\begin{eqnarray}
    x(\tau)&=&\int^\tau_{\tau_0} d\tau'/\theta
     = \frac{1}{1-p}\left[\frac{\tau}{\theta(\tau)}-
    \frac{\tau_0}{\theta_0}\right] \, , \quad p\ne 1
        \, , \label{x}
\end{eqnarray}
where $\theta_0=\theta(\tau_0)$. The marginal case $p=1$ was
studied in \cite{Gavin}.

The evolution of the parton gas can be conveniently studied by taking
moments of Eq.~(\ref{fint}) with respect to particle energies. Summing
over particle momenta gives a simple integral equation for the
particle density which has the simple solution
$n(\tau)=n_0\tau_0/\tau$ which also follows directly from
Eq.~(\ref{relax}). Multiplying by particle energy and summing
over momentum we obtain
\begin{eqnarray}
   e^{x}g(\tau) = h(\tau_0/\tau)
   +\int^x_0 dx' e^{x'}g(\tau')h(\tau'/\tau)  \, , \label{gint}
\end{eqnarray}
where
\begin{eqnarray}
   g(\tau) = \frac{\tau}{\tau_0}\frac{\epsilon(\tau)}{\epsilon(\tau_0)} \, ,
\end{eqnarray}
and
\begin{eqnarray}
    h(r)&=& \int^1_0d\cos(v)\sqrt{1+\cos^2(v)(r^2-1)}
        =\frac{1}{2}\left(r+\frac{\sin^{-1}\sqrt{1-r^2}}{\sqrt{1-r^2}}
        \right) \, . \label{h}
\end{eqnarray}
Here $v$ is the polar angle of the particle momenta with respect  to the
$z$-axis.  The function $h(r)$ is a monotonically increasing function between
$h(0)=\pi/4$ and $h(1)=1$. The function $g(x)$ is calculated numerically
and is shown in Fig.~\ref{fig2} for
various $\theta_0$ and $p<1$.

Performing a partial integration on Eq.~(\ref{gint}) gives
\begin{eqnarray}
  \int^x_0 dx' e^{x'} \frac{d}{dx'} \left[ g(\tau')h(\tau'/\tau) \right]
  =0   \, .\label{cond}
\end{eqnarray}
Note that by assuming an initially spherical symmetric distribution  we differ
slightly from Baym \cite{Baym}. He assumes the distribution  to be peaked in
transverse directions initially. The term $h(\tau_0/\tau)$ in  Eq.
(\ref{gint}) is then replaced by unity and the r.h.s. of  Eq.~(\ref{cond})
changes to $1-\pi/4$. The difference in initial  conditions will not affect
any of our subsequent arguments concerning  the $p$-dependence of the
thermalization.

When $p<1$, $x(\tau)$ increases with increasing $\tau$ as seen from
Eq.~(\ref{x}). The exponential factor $e^{x'}$ thus weights large $x'$
and Eq.~(\ref{cond}) implies for large $x$ or $\tau$
\begin{eqnarray}
   \frac{d}{d\tau'} \left[ g(\tau')h(\tau'/\tau) \right]_{|\tau'=\tau} = 0
   \, ,  \label{gcond}
\end{eqnarray}
since $d\tau=\theta\, dx$. The slope of $h(r)$ near $r=1$ is 1/3 and
so we conclude that $g(\tau)\propto \tau^{-1/3}$ for large $\tau$.
Thus $\epsilon(\tau)\propto \tau^{-4/3}$ which is the one-dimensional
hydrodynamical limit.

When $p>1$, $x(\tau)$ is negative and decreases with increasing $\tau$  to a
minimal but finite value $x_{\rm min}=-\tau_0/\theta_0/(1-p)$.  By
differentiating Eq.~(\ref{gint}) we find that $g'(\tau)$ is always  negative
and therefore $g(\tau)$ decreases monotonically. Yet, by  definition, $g$ is
positive and must therefore have an asymptotic  value larger than or equal to
zero.  From the integral  equation (\ref{gint}) we see that this value at
$\tau=\infty$ or $x=x_{\rm max}$ is non-vanishing and thus  $g(\infty)>0$.
Consequently, $\epsilon(\tau)=g(\infty) \tau_0/\tau$ at large times which is
the one-dimensional free streaming limit.

We conclude that thermalization will be reached when $p<1$ and the parton
gas will
expand hydrodynamically at large times whereas when $p>1$  it will continue to
stream freely. The marginal case $p=1$ was studied  by Gavin \cite{Gavin} who
found that the parton gas ends up in a state  between hydrodynamic expansion
and free streaming depending on  the size of the prefactor $\alpha$ where
$\theta=\alpha\tau$.   When $\alpha$ is small the collision time is always
relatively short  and the gas equilibrates near the hydrodynamic limit. When
$\alpha$  is large the collision time is always longer than the expansion time
and the gas continues to stream freely. One should, however, keep in  mind the
finite decoupling, freeze-out, or hadronization time. When  it is shorter than
the collision time, which will be the case for  large $\theta_0$ or $p$ close
to unity or larger, the parton gas does  not thermalize.

\section{Time Dependence of the Relaxation Time}

As we have just seen, the magnitude of the collision time as well as its
dependence on expansion time is crucial for the equilibration. We shall
therefore study the collision term and calculate it within perturbative QCD
where we know the scattering matrix elements between quarks and gluons. We
shall here only look at elastic scattering processes even though inelastic
processes has been found to be as important in both the case  of later parton
chemical equilibration \cite{Biro,LXES} and energy loss of a fast parton going
through a QGP\cite{GW}. Weldon has, however, shown \cite{Weldon} that
inclusion of absorption processes as well as production and absorption of
virtual particles reduces the energy loss further. Near chemical equilibrium
emission and  absorption should balance but the free streaming process drives
the system away from equilibrium and thus inelastic processes is a  possible
candidate responsible for further equilibration.

The collision integral for scattering particles from initial
states 1 and 2 to final states 3 and 4 is
\begin{eqnarray}
  \left(\frac{\partial f_1}{\partial t}\right)_{\rm coll}  &=&
  - (2\pi)^4
  \nu_2 \int d\Gamma_{{\bf p}_2}d\Gamma_{{\bf p}_3}d\Gamma_{{\bf p}_4}
  |M_{12\to 34}|^2
\nonumber\\ &\times&
  [ f_1f_2(1\pm f_3)(1\pm f_4) - f_3f_4(1\pm f_1)(1\pm f_2)]
    \delta^4(p_1+p_2-p_3-p_4)\, ,  \label{coll}
\end{eqnarray}
where $p_i$ are the parton four-momentum. We assume they are
massless, {\it i.e.}, $E_i=|{\bf p}_i|$. The $(1\pm f_i)$ factors
correspond physically to the Pauli blocking of final states, in the
case of fermions, and to (induced or) stimulated emission, in the case
of bosons. $\nu_2$ is the statistical factor, $16$ for gluons and $12N_f$
for quarks and antiquarks. $|M_{12\to 34}|^2=|{\cal M}_{12\to 34}|^2/
(16E_1 E_2 E_3 E_4)$ is the matrix element squared summed over final
states and averaged over initial states. For scattering of gluons
\begin{eqnarray}
   |{\cal M}^{(gg)}_{12\to 34}|^2 = \frac{9}{2}  g^4 \,
  \left( 3-\frac{us}{t^2}-\frac{st}{u^2}
                      -\frac{ut}{s^2} \right)  \, ; \label{Mgg}
\end{eqnarray}
quark-gluon and quark-quark interactions are just $4/9$ and $(4/9)^2$
times weaker respectively near forward scattering ($t\simeq0$). In a
$t=\omega^2-q^2$ channel, a singularity occures for small momentum
${\bf q}$ and energy $\omega$ transfers. In a medium the $t^{-2}$
singularity is screened as given by Dyson's equation  in which a
gluon self-energy $\Pi_{L,T}$ is added to the propagator
\begin{eqnarray}
    t^{-1} \to \omega^2-q^2-\Pi_{L,T} \, . \label{Dyson}
\end{eqnarray}
As was shown in \cite{BP1,BP2,eta} that Debye screening and dynamical
screening due to Landau damping effectively screen the longitudinal
and transverse interactions off in most transport problems at a length
scale of order the Debye screening length $q_D^{-1}$.  For small
momentum transfer, $q \ll E_1,E_2$, one can split the matrix element
into longitudinal and transverse parts \cite{eta},
\begin{equation}
  |M_{gg}|^2 = \frac{9}{8} g^4 \left[\frac{1}{q^2+\Pi_L}
  -\frac{(1-\omega^2/q^2)\cos\phi}
  {q^2-\omega^2+\Pi_T}\right]^2\, ,
  \label{longt}
\end{equation}
where $\cos\phi=({\bf v}_1\times\hat{\bf q})\cdot({\bf v}_1\times\hat{\bf
q})$.
The gluon self energies, $\Pi_L$ and $\Pi_T$ are given in the
long wavelength limit ( $q \ll T$) by \cite{Weldon2}
\begin{eqnarray}
  \Pi_L(q,\omega)&=&q_D^2\left[1-\frac{\omega}{2q}
  \ln\left(\frac{q+\omega}{q-\omega}\right)\right] \, ,\\
  \Pi_T(q,\omega)&=&q_D^2\left[\frac{\omega^2}{2q^2}
  +\frac{\omega}{4q}(1-\frac{\omega^2}{q^2})
  \ln\left(\frac{q+\omega}{q-\omega}\right)\right] \, .
\end{eqnarray}
The Debye screening wavenumber in thermal QCD is
$q_D^2=g^2(1+N_f/6)T^2$ where $N_f$ is the number of quark flavors.

The Boltzmann equation with the full collision term, Eq.~(\ref{coll}),
has been solved for quark-gluon plasmas near equilibrium and a number of
transport coefficients have been calculated to leading orders in the coupling
constant \cite{BP1,BP2,color}. For viscous and thermal relaxation as well as
momentum stopping, the ``transport relaxation time" is generally
\begin{eqnarray}
   \tau_{\rm tr} \simeq \left(\alpha_s^2 \ln (\frac{1}{\alpha_s})
   \lambda T\right)^{-1}
  \, , \label{tr}
\end{eqnarray}
where $\lambda$ (the ``fugacity") is the ratio of the actual density to the
one in chemical equilibrium at temperature $T$. This relaxation time may be
used at later times when the parton gas  is near equilibrium. In Bjorken flow
the temperature scales like  $T\propto\tau^{-1/3}$, {\it i.e.},
$\tau_{\rm tr}\propto\tau^{1/3}$.  Since this power is less than unity,
the parton
gas should thus  equilibrate according to the analysis in the previous section.

In nuclear collisions the parton gas may be far from equilibrium  when first
produced and the expansion may also drive it out of  equilibrium as in free
streaming. Solving the Boltzmann equation  thus becomes a very difficult
non-linear problem that requires  major computational efforts which is being
undertaken in a number of parton cascade models as, {\it e.g.}, in
\cite{Geiger}. We take another approach in this paper. As mentioned above,
hydrodynamics does not apply at early times because of long  viscous
relaxation times and the parton gas is expected to expand  as free streaming
initially. With this initial ansatz in Eq.~(\ref{fstream})  for the
distribution function we can calculate the change  in the distribution
function at early times from the Boltzmann equation with the full collision
term of Eq.~(\ref{coll}). More specifically, from the entropy density
\begin{eqnarray}
   s(\tau) = - \sum_{\bf p} [f\ln f \mp(1\pm f)\ln(1\pm f)] \, ,\label{s}
\end{eqnarray}
we have calculated the entropy production
\begin{eqnarray}
    (\frac{\partial s}{\partial t})_{\rm coll} = \sum_{\bf p}
   (\frac{\partial f}{\partial t})_{\rm coll} \ln\left(\frac{f}{1\pm f}\right)
                  \, ,\label{ds}
\end{eqnarray}
in the free streaming phase
by approximating $f$ by $f_0$ [see Eq. (\ref{f0}) in the Appendix].
The initial entropy production can be estimated
analytically with the full collision term to leading logarithmic
order in the coupling constants and details of this calculation is
given in the Appendix. Ignoring quarks the final result is
[see (\ref{sfree2})]
\begin{eqnarray}
    \left(\frac{\partial s}{\partial t}\right)_{\rm coll} =
    \nu_g^2\frac{9}{8\pi^4}\alpha_s^2 T_0^4
 \lambda_{0,g}^2
    \ln\left[\frac{9\tau}{\pi\lambda_{0,g}\alpha_s\tau_0}\right]
    \ln(\frac{2\tau}{\tau_0})\, , \label{sfree1}
\end{eqnarray}
where $\lambda_{0,g}=\exp(\mu_{0,g}/T_0)$ is the ratio of the initial density
to that in chemical equilibrium.

We can match this entropy production to that obtained in  the relaxation time
approximation thus determining an efective and momentum averaged relaxation
time $\theta$.  The relaxation time and entropy production during this early
period, $\tau_0 \ll \tau \ll \theta$, are different from  later times
$\tau\raisebox{-.5ex}{$\stackrel{>}{\sim}$}\theta$ when collisions change the
free streaming distribution
functions. The entropy production for the initial free streaming
in the relaxation time approximation is
from Eq.~(\ref{ds})
\begin{eqnarray}
 \left(\frac{\partial s}{\partial t}\right)^{\rm relax}_{\rm coll}  &=&\,
 - \int d\Gamma_{\bf p} \left(\frac{f_0-f_{\rm eq}}{\theta}\right)
  \frac{\sqrt{p_\perp^2+(p_z\tau/\tau_0)^2} -\mu_0}{T_0} \nonumber\\
  &=& \frac{\epsilon(T_0)}{\theta T_0}
    \left\{\frac{1}{4}\left[1+\frac{\tau_0^2}{\tau^2}
    \frac{\ln(1+\sqrt{1-\tau_0^2/\tau^2}\,)}{\sqrt{1-\tau_0^2/\tau^2}}\right]
    \left[\frac{\tau_0}{\tau}+\frac{\sin^{-1}(\sqrt{1-\tau_0^2/\tau^2})}
    {\sqrt{1-\tau_0^2/\tau^2}}\right]-\frac{\tau_0}{\tau}\right\} \nonumber\\
    &\simeq& \frac{\pi}{8}\frac{\epsilon(T_0)/T_0}{\theta}
    \, ,\quad \tau\gg\tau_0 \, . \label{sinit}
\end{eqnarray}
As in Eq.~(\ref{sfree1}) we ignore quarks;
$\epsilon(T_0)\approx \nu_g(\pi^2/30)T_0^4\lambda_{0,g}$ is the
initial gluon energy density.
By equating the entropy production of Eq. (\ref{sinit}) for
$\tau\gg\tau_0$ to that
from the perturbative QCD collision term Eq.~(\ref{sfree1}) we find
\begin{eqnarray}
 \frac{1}{\theta} \simeq 1.43 \alpha_s^2 T_0 \lambda_{0,g}
    \ln\left[\frac{3\pi\tau}{2\lambda_{0,g}\alpha_s\tau_0}\right]
    \ln(\frac{2\tau}{\tau_0}) \, . \label{theta}
\end{eqnarray}
Note that it is the initial temperature $T_0$ that enters here and  not $T(t)$
as in Eq.~(\ref{tr}). The collision time depends only  logarithmically on the
expansion time. As explained in the Appendix, the relaxation time is only
weakly time dependent  in a free-streaming parton gas because  the phase space
for small momentum scattering opens up quadratically  with time thus
effectively compensating the decrease in parton densities. On the other hand,
if large momentum transfers are imposed to each parton scattering, the entropy
production rate will decrease quadratically with time, leading to a much
stronger time dependence of the relaxation time. The long range  interactions
(small momentum transfers) are therefore very important in an expanding parton
gas. The two logarithms  in Eq.~(\ref{theta}) arise from integrals over
momentum and  energy  transfers respectively. The ``fugacity"  factor
$\lambda_0$ arises from the correspondingly  smaller density of scatterers.

To estimate the relaxation time $\theta$, we need to know the  initial
condition $\tau_0$, $T_0$, $\lambda_0=e^{\mu_0/T_0}$ which so far can only be
determined from model calculations. From the HIJING model calculation
\cite{Wang,HIJING,PLXW} it was found  that at $\tau_0=0.7$ fm/$c$ the produced
partons in central $Au+Au$ collision at RHIC energy can reach a local isotropy
in  momentum distribution temporarily with effective temperature $T_0=0.57$
GeV; in addition $\lambda_{0,g}=0.09$ and $\lambda_{0,q}=0.02$.  At LHC
energies $\tau_0=0.5$ fm/$c$, $T_0=0.83$ GeV, $\lambda_{0,g}=0.14$ and
$\lambda_{0,q}=0.03$. Eskola {\it et al.} \cite{Kari} found similar results
for the temperature $T_0= 1 (1.5)$ GeV at the RHIC (LHC) energy in their
minijet plasma calculation. However, they used a smaller initial time,
$\tau_0=0.1$ fm/$c$, and find $\lambda_{0,g}\sim 1$ and $\lambda_{0,q}\sim 0$
If one were to allow this minijet plasma to stream freely  shortly after
$\tau_0$ the result would be consistent with HIJING estimates at the later
time $\tau_0\simeq 0.7$ fm/c. The newest set of parton distribution functions
in the calculation of Eskola et al. also increase the initial parton density,
especially at LHC. In the parton cascade model \cite{Geiger} the initial
parton density is found to be larger due to a different treatment of soft
parton interactions. These numbers for $\tau_0$ and $T_0$ are surprisingly
similar to those found by Shuryak \cite{Shuryak} in a different analysis. By
estimating the particle rapidity distributions $dN/dy$ in relativistic nuclear
collisions he obtains a particle density in the Bjorken scenario $(dN/dy)/(\pi
R^2\tau)$ which at a time $\tau=\tau_{\rm coll}\sim (\alpha_s T)^{-1}$ is
assumed to be the same as the equilibrium one $\sim T^3$. Hereby the initial
values for $T$ and $\tau$ are found. One should bear in mind that all these
estimates are based on perturbative QCD inspired models. There are many
uncertainties (see, e.g., \cite{wangqm95}) due to our limited knowledge of
strong interactions.

If we ignore the slow logarithmic time dependence and assume
$T_0\simeq 1/\tau_0$ and $\lambda_{0,q}\sim 0$,
as motivated by the above mentioned models, we find
\begin{eqnarray}
    \theta_0 \simeq \alpha_s^{-2} \lambda_{0,g}^{-1}\tau_0 \, .  \label{tappr}
\end{eqnarray}
With $\alpha_s\simeq 0.3$ we find a rather long collision time
even if $\lambda_{0,g}=1$.

\section{Entropy Production}

With the weak time dependence of the relaxation time, we can now estimate the
entropy production during the early thermalization. We still assume the
relaxation time has a power dependence on time as in Eq.~(\ref{power}) with
small $p$. The scaling behavior of the function $g(\tau)$ with
$\theta_0=\theta(\tau=\tau_0)$ can be obtained when  $\theta_0\gg\tau_0$. In
that case the integral equation (\ref{gint})  for $g(\tau(x))$ depends almost
solely on $x$, since  $x'/x\simeq (\tau'/\tau)^{1-p}$. Thus for given $p$,
$g(\tau(x))$  is a generic function of $x$. Its behavior at large $x$ is
\begin{eqnarray}
    g(\tau(x)) &\simeq& g_p x^{-1/3(1-p)}
           = g_p \left[ (1-p)\frac{\theta_0}{\tau_0}\right]^{1/3(1-p)}
             \left(\frac{\tau}{\tau_0}\right)^{-1/3} \ ,    \label{gapp}
\end{eqnarray}
where $g_p$ is some $p$-dependent constant of order unity.
In the case $p=0$ Baym \cite{Baym} found $g_p=1.22$ but due
to the different initial conditions (isotropic versus peaked in the
transverse directions) this value differ from
our case by a factor $\pi/4$ when $\theta_0\gg\tau_0$.

Until the parton gas reaches equilibrium its entropy at time $\tau$
is always less than the equilibrium entropy at temperature $T(\tau)$,
{\it i.e.}, $s\le s_{eq}=(4/3)\epsilon/T=s_0g(\tau)T_0\tau_0/(T\tau)$.
Thus, the total entropy is
\begin{eqnarray}
   S\le S_0\,g(\tau) \frac{T_0}{T}
   =S_0\,\left(\frac{\tau}{\tau_0}\right)^{1/4} g(\tau)^{3/4}
   \, . \label{sbaym}
\end{eqnarray}
If the parton gas equilibrates the equal sign holds at large
times and the final entropy is  from Eqs. (\ref{gapp},\ref{sbaym})
\begin{eqnarray}
    \frac{S_f}{S_0} = s_p (\theta_0/\tau_0)^{1/4(1-p)} \, , \label{sparam}
\end{eqnarray}
where $s_p$ is a $p$-dependent number.
In Fig.~\ref{fig3}, the final entropy is plotted for various
values of $p$ and $\theta_0$. The formula Eq.~(\ref{sparam}) is
a good approximation with coefficient $s_p\sim 1$. For comparison
the final entropy calculated by Baym
\begin{eqnarray}
    \frac{S_f}{S_0} = 1.16\,
   (\theta_0/\tau_0)^{1/4} \, .  \label{sbaymf}
\end{eqnarray}
with a constant relaxation time ($\theta=\theta_0$, $p=0$) is also
shown in Fig.~\ref{fig3}, being multiplied by a
factor $(\pi/4)^{3/4}$ because of the different initial conditions.

It is interesting to compare the entropy production to the case when the
parton gas is streaming freely until a time $\tau_{\rm sudden}$
when the gas suddenly collides
violently and immediately thermalizes.
Conserving energy per volume (but not particle density) we find
\begin{eqnarray}
    \frac{S_f}{S_0} = (\frac{\pi}{4})^{3/4}
   (\tau_{\rm sudden}/\tau_0)^{1/4} \, ,
     \label{ssudden}
\end{eqnarray}
which is only slightly lower than Eq.~(\ref{sbaymf}) if
$\theta_0\simeq\tau_{\rm sudden}$. The exact way by which the parton
gas equilibrates is therefore not so important; it is the collision
time that determines the entropy production.

{}From Eq.~(\ref{sbaymf},\ref{sparam}) it is evident that the final
entropy production increases slowly with the collision time. Yet
it takes much longer time and at a given time
($\tau\raisebox{-.5ex}{$\stackrel{<}{\sim}$}\theta$)
the entropy production rate [cf. Eq.~(\ref{sinit})] is inversely
proportional to $\theta$. If the parton gas decouples, freeze out or
hadronize at  time, $\tau_d$, and entropy is no longer produced,
then the total entropy production  will decrease as $\theta$
increases above $\tau_d$.

With the approximate collision time of Eq.~(\ref{tappr})
and assuming $T_0\simeq 1/\tau_0$ the final entropy is obtained from
Eq.~(\ref{sparam}) or (\ref{sbaymf})
\begin{eqnarray}
   \frac{S_f}{S_0} \simeq \alpha_s^{-1/2} \lambda_{0,g}^{-1/4} \, .
\end{eqnarray}
For $\alpha_s\sim 0.3$ this gives an
increase in entropy by a factor of 2-3 when varying $\lambda_0$
from unity down to 0.1.

\section{Summary}

We have studied a one-dimensionally expanding parton gas created in the wake
of nuclear collisions. Within the Boltzmann equation in the relaxation time
approximation we find that the rapid expansion is closer to free streaming
than hydrodynamic expansion for times shorter than typical collision times of
partons, $\tau_0\raisebox{-.5ex}{$\stackrel{<}{\sim}$}\tau
\raisebox{-.5ex}{$\stackrel{<}{\sim}$}\theta$.
Only at times much larger than the
characteristic collision times,$\tau\gg\theta$ may the parton gas
thermalize and expand hydrodynamically. However,
if the collision time increases with time the gas may never thermalize.
Parametrizing the collision time  as $\theta=\theta_0(\tau/\tau_0)^p$
the condition for equilibration and hydrodynamical expansion is $p<1$. We
calculate how much entropy is produced in the collisions. These
different expansion modes and the corresponding parton phase
space distributions at different stages of evolution in high-energy
heavy ion collisions are illustrated in Fig.~\ref{fig4}.

We have described the parton scatterings by perturbative QCD in order to
estimate the important collision time and its dependence on expansion time.
The effective ``out of equilibrium'' collision time differ from the standard
transport relaxation time,
$1/\tau_{\rm tr}\simeq \alpha_s^2\ln(1/\alpha_s)T(\tau)$, with a weaker
time dependence. However, it is still the Debye screening and
Landau damping that  screen the singular forward scattering processes.

We find that the parton gas does equilibrate eventually with these  collision
times but only after having undergone a phase of free streaming and gradual
thermalization where considerable entropy is produced (``after-burning"). The
final entropy and thus particle density depends on the collision time as well
as the initial conditions (a ``memory effect"). For
various models predicting the preequilibrium scenarios
the entropy production is significant.
The total entropy and particle production is estimated to be
doubled or tripled with respect to the initial value.

These estimates do not include particle production which by itself
adds to the entropy production. On the other hand particle production
will also increase the density and thus shorten the effective
collision time which leads to a decrease in entropy production
according to Eq.~(\ref{sparam}).

Most analyses assume a constant density in space but large density
fluctuations may well be present in the initial parton plasma. This
will increase the average entropy production  for both elastic and
inelastic scatterings since these are proportional to the initial
densities  squared as well as the final densities through the
stimulated emission factors (for bosons) or Pauli blocking
factors (for fermions). High density regions (``hot spots") will
equilibrate thermally and chemically faster than low density regions.
At the same time, however, the free streaming will tend to reduce
density fluctuations.

It was emphasized in the Appendix that the very singular  small momentum
transfers provides strong scattering and the opening up of phase space
compensates for the decreasing densities. If a larger momentum
transfer cut-off of the order of particle momenta ($\sim$ temperature)
is applied then the collision time will increase quadratically with
expansion time and the parton gas will never thermalize. Also, when
the longitudinal extension of the system exceeds the transverse size,
which is in the order of the nuclear radius, the expansion proceeds
in three dimensions. The densities will then decrease cubically with
expansion time and collision might never catch up with the expansion.

\section*{Acknowledgement}

This work was supported by the Director, Office of Energy Research,
Office of High Energy and Nuclear Physics, Division of Nuclear Physics,
of the U.S. Department of Energy under Contract DE-AC03-76SF00098,
and the Danish Natural Science Research Council. Discussions with
G.~Baym, K.~J.~Eskola and P.~Siemens are
gratefully acknowledged.

\appendix

\section*{The relaxation time}

In this appendix we give a detailed derivation of the entropy
production by elastic parton collisions in a system near
the free streaming case at times $\tau_0\ll\tau\ll\theta$.

The initial distribution of partons has been estimated in several models
\cite{Geiger,HIJING}. Typically, one finds that the partons are formed within
the first  $\tau_0\simeq 0.2-0.7$ fm/$c$ after the nuclear collision and that
rapid  longitudinal expansion takes place. The local momentum distribution
of partons at the formation time is not isotropic but forward/backward
peaked, {\it i.e.}, $\langle|p_z|\rangle\gg\langle p_\perp\rangle$.
However, due to streaming the local distribution changes rapidly
since $\langle|p_z|\rangle$ scales as $\tau_0/\tau$ [see Eq.~(\ref{fstream})]
and at later times the particles have
$\langle|p_z|\rangle\ll\langle p_\perp\rangle$. At the cross over time
$\tau_0$ the parton gas is isotropic in momentum space and seems to be in
approximate  thermal equilibrium with temperature $T_0$ even though it is
streaming freely in space and time.  From Eq.~(\ref{fstream}) we obtain the
free streaming  distribution function
\begin{eqnarray}
    f_0({\bf p},\tau) =
       \left[\exp(\frac{\sqrt{p_\perp^2+(p_z\tau/\tau_0)^2}
               -\mu_0}{T_0})\pm 1\right]^{-1} \, ,  \nonumber\\   \label{f0}
\end{eqnarray}
where $\mu_0$ is the chemical potential determined by the density. Since the
parton gas may not have reached chemical equilibrium yet, $\mu_0$ may not
vanish. In fact most models predict that particle densities are rather low
initially \cite{Geiger,PLXW,Kari} and that $-\mu_0\simeq (1-2)T_0$. The
temperature increases with collision energy and typically $T_0\simeq 0.5-2$GeV.
We shall use Eq.~(\ref{fstream}) with Eq.~(\ref{f0}) for the free
streaming initially with parameters $\tau_0$, $T_0$ and $\mu_0$.
The entropy production due to collisions is negligible at times
around $\tau_0$ because the parton gas is near thermal equilibrium
and so we shall ignore collisions earlier than $\tau_0$. On
the other hand, continuing particle production will produce
entropy but we shall not consider that contribution here.

With the free streaming distribution function, we obtain from the
Boltzmann equation by changing variables from
$p_{z,i}\to p_{z,i}\tau_0/\tau$, {\it i=}1,2, and
$q_z\to q_z\tau_0/\tau$
\begin{eqnarray}
 \left(\frac{\partial s}{\partial t}\right)_{\rm coll}  &=&\,
     - 2\pi\nu_1\nu_2   \frac{\tau_0^3}{\tau^3}
 \int d\Gamma_{{\bf p}_1}d\Gamma_{{\bf p}_2}d\Gamma_{{\bf q}}
|M_{12\to 34}({\bf q}_\perp,q_z\frac{\tau_0}{\tau})|^2
  \,  f_0(E_1)f_0(E_2)
  (1\pm f_0(E_3))(1\pm f_0(E_4))
  \nonumber\\ &\times& \frac{E_1-\mu_0}{T_0}
     [1-\exp(\frac{E_1+E_2-E_3-E_4}{T_0} ) ]
   \delta (\tilde{E}_1+\tilde{E}_2-\tilde{E}_3-\tilde{E}_4) \, ,
     \label{ds2}
\end{eqnarray}
where $\tilde{E}_3=\sqrt{({\bf p}_{1\perp}+{\bf q}_\perp)^2+
(p_{1,z}+q_z)^2\tau_0^2/\tau^2}$
and $\tilde{E}_1=\sqrt{p_{1\perp}^2+(p_z\tau_0/\tau)^2}$.
The same expressions are valid for {\it i=}2, 4
when the sign of ${\bf q}$ is changed.

At this point we want to emphasize the importance of small momentum
transfer processes, $q\sim q_D\sim gT$ as compared to large momentum
transfer ones, $q\sim T$. For the latter the exponential in Eq.~(\ref{ds2})
can be ignored because $E_4+E_3-E_2-E_1 \sim T$. One then
finds that the integrations over the energy conservation $\delta$-function
removes a factor $\tau_0/\tau$ leaving an entropy production rate decreasing
quadratically in time. This just reflects that the particle densities of
the two scatterers decrease as $\tau_0/\tau$.
The small momentum transfers have, however, a very singular scattering matrix
element and, as we will now show, the phase space opens up quadratically
with time for small $q$ - effectively {\em compensating} the decreasing
densities of scatterers.

Expanding around small $q$ to second order, the term in the square
bracket representing the difference between scattering in and out is
\begin{eqnarray}
 1-\exp(\frac{E_1+E_2-E_3-E_4}{T_0})&\simeq&
 \frac{({\bf v}_1-{\bf v}_2)\cdot{\bf q}}{T_0}
 -\frac{[({\bf v}_1-{\bf v}_2)\cdot{\bf q}]^2}{2T_0^2}
\nonumber \\
 &+& \frac{q^2-({\bf v}_1\cdot{\bf q})^2}{2E_1T_0} +
 \frac{q^2-({\bf v}_2\cdot{\bf q})^2}{2E_2T_0}  \, . \label{sqrbrac}
\end{eqnarray}
Only when $\tau=\tau_0$ is $\tilde{E}=E$
and energy conservation requires that Eq.~(\ref{sqrbrac}) vanishes.
Due to symmetry the first term vanishes when integrated over ${\bf p}_1$
and ${\bf p}_2$ and the leading term is second order in $q^2$.

For small $q$ we can also replace $E_3$ by $E_1$ and $E_4$ by $E_2$
in the distribution functions. Thus we have
\begin{eqnarray}
 \left(\frac{\partial s}{\partial t}\right)_{\rm coll}  &=&\,
     - 2\pi\nu_1\nu_2   \frac{\tau_0^3}{\tau^3}
  \int d\Gamma_{{\bf p}_2} f_0(E_2)(1\pm f_0(E_2))
 \int d\Gamma_{{\bf p}_1}\frac{E_1-\mu_0}{T_0}f_0(E_1)(1\pm f_0(E_1))
          \nonumber\\
 &\times& \int d\Gamma_{\bf q}
    |M_{12\to 34}({\bf q}_\perp,q_z\frac{\tau_0}{\tau})|^2
    \int d\omega \delta(\omega-\tilde{E}_1 +\tilde{E}_3)
    \delta(\omega+\tilde{E}_2 -\tilde{E}_4 ) \nonumber \\
    &\times&
    \left\{\frac{q^2-({\bf v}_1\cdot{\bf q})^2}{2E_1T_0} +
 \frac{q^2-({\bf v}_2\cdot{\bf q})^2}{2E_2T_0}
 -\frac{[({\bf v}_1-{\bf v}_2)\cdot{\bf q}]^2}{2T_0^2}\right\}.  \label{ds3}
\end{eqnarray}
Here we have introduced an auxiliary integral over energy
transfer $\omega$.

We can use up these $\delta$-functions by performing the angular
integrals $d\Omega_i$, $i=$1,2. For example,
\begin{eqnarray}
   I_1 &\equiv&
   \int d\Omega_1 \delta(\omega-\tilde{E}_1 +\tilde{E}_3)
    = \int_0^{2\pi} d\phi_1\int_0^{\pi} \sin\theta_1 d\theta_1
     \nonumber\\ &\times&
     \delta\left[ \omega -
     \frac{\cos\phi_1\sin\theta_1 q_\perp+\cos\theta_1 q_z\tau_0^2/\tau^2}
     {\sqrt{1-\cos^2\theta_1(1-\tau_0^2/\tau^2)}} \right]     , \label{do}
\end{eqnarray}
where $\theta_1$ and $\phi_1$ are the polar and azimuthal angles of
${\bf p}_1$ in a coordinate system with $z$-axis along the collision
beam direction. For $\tau=\tau_0$ the prefactor in Eq.~(\ref{sqrbrac})
vanishes due to energy conservation. For $\tau\gg\tau_0$ we can ignore
the $(q_z\tau_0/\tau)^2$ term in Eq.~(\ref{do}), which then yields
\begin{eqnarray}
   I_1  &\simeq &
   \frac{2}{q_\perp}\int_{-1}^{1}dx\int_{-1}^{1} \frac{dy}{\sqrt{1-y^2}}
     \delta\left( \frac{\omega}{q_\perp} -\frac{y\sqrt{1-x^2}
    }{\sqrt{1-bx^2}}  \right)     \, ,
   \nonumber \\ && \label{d1}
\end{eqnarray}
where $b=1-\tau_0^2/\tau^2$, $x=\cos\theta_1$ and $y=\cos\phi_1$.
Changing variables to $\sin\chi=x
\sqrt{(1-\omega^2/q_\perp^2)/(1-b\omega^2/q_\perp^2)}$,
this integral gives
\begin{eqnarray}
 I_1&=& \frac{4}{q_\perp\sqrt{1-b\omega^2/q_\perp^2}}
   \int_0^{\pi/2} d\chi \sqrt{1-x^2}
   \nonumber \\
   &\simeq& \frac{4}{q_\perp\sqrt{1-b\omega^2/q_\perp^2}}
    \, , \, {\rm for}\, \tau_0/\tau \ll 1 \, . \label{d2}
\end{eqnarray}

The angular integrals of $d\Omega_2$ yields the same factor
as Eq.~(\ref{d2}) and the integral over energy transfers in
Eq.~(\ref{ds3}) thus gives at large times
\begin{eqnarray}
  \int d\omega I_1I_2 &=& \int d\omega d\Omega_1 d\Omega_2
   \delta(\omega-\tilde{E}_1 +\tilde{E}_3)
   \delta(\omega+\tilde{E}_2 -\tilde{E}_4 )
  \nonumber\\
   &\simeq& 16 \int_{-q_\perp}^{q_\perp} \frac{d\omega }{
   q_\perp^2 -b\omega^2 }
   \simeq \frac{32}{q_\perp} \ln \left( \frac{2\tau}{\tau_0}\right) \, .
     \label{d3}
\end{eqnarray}

An angular dependence arising from Eq.~(\ref{sqrbrac}) should also
be included when performing integration in Eq.~(\ref{do}).
For example, an extra factor of $\cos^2\theta_1$ leads to an
additional factor of 1/3 in Eq.~(\ref{d3}). In addition the
matrix element depends on energy transfer through the
transverse part of the self-energy in Eq.~(\ref{longt}).
After evaluating the integrals over momentum transfers this
dependence is, however, only logarithmic in $\omega$ \cite{eta}
and can be ignored.

Let us first consider the entropy production due to gluon-gluon
scatterings and include quarks later.
We assume for convenience that
$\mu_0\raisebox{-.5ex}{$\stackrel{<}{\sim}$} -T_0$
which allows us to use Boltzmann distribution
functions, $f_0(E)=\exp((\mu_0-E)/T_0)=\lambda_0\exp(-E/T)$.
The momentum integrals of ${\bf p}_1$ and ${\bf p}_2$ in Eq.~(\ref{ds3})
are straight forward, which leave one remaining integral
over momentum transfer,
\begin{equation}
 \left(\frac{\partial s}{\partial t}\right)_{\rm coll} =
 \frac{3\nu_g^2}{2\pi^6} \alpha_s^2   \frac{\tau_0^3}{\tau^3}
  \ln(\frac{2\tau}{\tau_0}) T_0^4 \lambda_{0,g}^2
 \frac{3}{2} \int d^3q \frac{q^2}
    {q_\perp[q_\perp^2+(q_z\tau_0/\tau)^2+q_D^2]^2} \, , \label{sqint}
\end{equation}
where, we have for simplicity approximated  $\Pi_L$ by $q_D^2$ in the matrix
element of Eq. (\ref{Mgg}). Furthermore, we have replaced the transverse part
of interaction by the longitudinal one times a factor $\langle
\cos^2\phi\rangle=1/2$. Both these approximations are exact to leading
logarithmic order in the coupling constant for the calculation of a number of
transport coefficient \cite{eta}. Note that the limit for the integral  over
$q_z$ is $\pm q_{\rm max}\tau/\tau_0$ and the maximum momentum transfer,
$q_{\rm max}$, is determined by the distribution functions which cut off large
momentum transfers. It has been estimated in \cite{eta} to be $q_{\rm max}\sim
3T_0$. The integral over momentum transfers in (\ref{sqint}) then gives
$(\tau/\tau_0)^3\pi^2\ln(2q_{max}/q_D)$ when $\tau/\tau_0\ll q_{\rm max}/q_D$.
It includes the usual logarithm (see, {\it e.g.}, \cite{eta} for details)
as well as a
factor $(\tau/\tau_0)^3$ from the integral over $q_z$.  The entropy production
in QCD to leading order in the coupling coupling constant is thus
\begin{eqnarray}
    \left(\frac{\partial s}{\partial t}\right)_{\rm coll} =
    \frac{9}{8\pi^4}T_0^4 \alpha_s^2
    \left[ \nu_g^2\lambda_{0,g}^2
     + \frac{4}{9}\nu_g\nu_q\lambda_{0,g}\lambda_{0,q}
     + (\frac{4}{9})^2\nu_q^2\lambda_{0,q}^2 \right]
    \ln\left[\frac{4 q_{\rm max}^2}{q_D^2} \right]
    \ln(\frac{2\tau}{\tau_0})\, . \label{sfree}
\end{eqnarray}
Here, we have included contributions from quarks and antiquarks
to the entropy production; $\nu_g=16$, $\nu_q=12N_f$,
$\lambda_q=e^{\mu_q/T}$ and we assume $\lambda_q=\lambda_{\bar{q}}$.
The quark-gluon and quark-quark forward scattering interactions
are smaller than the gluon-gluon ones
by a factor (4/9) and $(4/9)^2$ respectively.
According to the models in Refs.~\cite{Geiger,Wang,Biro}
fewer quark and antiquark than gluons are produced in relativistic nuclear
collisions, {\it i.e.}, $\lambda_{0,q\bar{q}}\ll\lambda_{0,g}$.

The Debye screening mass in a quark-gluon gas in thermal and chemical
equilibrium with no net baryon density ({\it i.e.},
$\mu_g=\mu_q=\mu_{\bar{q}}=0$) is $q_D^2 = 4\pi(1+N_f/6)\alpha_sT^2$. Out of
chemical equilibrium when $\mu_q=\mu_{\bar{q}}$ and $\mu_{g,q}\ll -T$ the Bose
and Fermi distribution functions can be replaced by Maxwell-Boltzmann
distribution functions and we find
\begin{eqnarray}
 q_D^2 = \frac{24}{\pi}(\lambda_g+\frac{N_f}{3}\lambda_q)\alpha_sT^2 \, .
\end{eqnarray}
The quark and gluon densities and thus fugacities decrease with time due to
the one-dimensional expansion such that $\lambda=\lambda_0 \tau_0/\tau$.

Inserting these expressions for $q_{\rm max}$ and $q_D$ in Eq.~(\ref{sfree})
we find the entropy production rate
\begin{eqnarray}
    \left(\frac{\partial s}{\partial t}\right)_{\rm coll} &=&
    \frac{9}{8\pi^4}\alpha_s^2 T_0^4
    \left[ \nu_g^2\lambda_{0,g}^2
     + \frac{4}{9}\nu_g\nu_q\lambda_{0,g}\lambda_{0,q}
     + (\frac{4}{9})^2\nu_q^2\lambda_{0,q}^2 \right]
 \nonumber\\ &\times&
    \ln\left[\frac{3\pi\tau}{2
(\lambda_{0,g}+\frac{N_f}{3}\lambda_{0,q})  \alpha_s\tau_0}\right]
    \ln(\frac{2\tau}{\tau_0})\, ,
 \nonumber\\ &&\label{sfree2}
\end{eqnarray}
to leading order in a free streaming gluon gas.
The result is approximately  valid for
$\tau\raisebox{-.5ex}{$\stackrel{>}{\sim}$} 2\tau_0$ and as long
as the free streaming assumption is valid, {\it i.e.},
$\tau\raisebox{-.5ex}{$\stackrel{<}{\sim}$}\theta$.

We emphasize the important result that $\tau^3$ factor is cancelled by the
integral over momentum transfers. Only a slow logarithmic dependence on time
remains. The physical explanation for this cancellation is the following. From
Eq.~(\ref{ds2}) a factor $(\tau_0/\tau)^2$  appears from the substitutions of
integration variables $p_{z,i}, i=$1,2. This represents the fact that the
densities of each of the colliding partons drop like $\tau_0/\tau$. If we
keep the original momentum transfer variables ({\it i.e.}, do not replace
$q_z\to q_z\tau_0/\tau$ in Eq.~(\ref{ds2})) the factor in the square bracket,
Eq.~(\ref{sqrbrac}), leads to a factor $(q\tau/\tau_0)^2$. This is because
the phase space for small momentum scattering {\it opens up quadratically
with time} and it {\it balances the decrease in parton densities}. On the
other hand, for large momentum transfers the exponential in the square
bracket, Eq.~(\ref{sqrbrac}), simply vanish leaving a factor of unity which
leads to a much reduced entropy production rate decreasing quadratically with
time. The long range interactions (small momentum transfers) are therefore
very important in expanding plasmas and sensitive to screening or the cutoff
as is applied in some models \cite{Geiger}.

\begin{figure}
\caption{Total entropy in one-dimensional
viscous hydrodynamical expansion of Eq.~(\protect\ref{S})
for various viscosities $\eta=\tilde{\eta}T^3$.  For comparison,
the upper limit on the entropy by solving the Boltzmann equation,
Eq.~(\protect\ref{sbaym}), is shown with dashed
curve for $\theta=\tau_\eta$ in the case $\tilde{\eta}=16$.
In both cases $T_0=\protect\hbar/\tau_0$ is assumed.}
\label{fig1}
\end{figure}

\begin{figure}
\caption{Qualitative picture of the expanding system following a
nuclear collision. The formation
time from $0-\tau_0$, free streaming from $\tau_0-\theta$, and
hydrodynamical flow for $\tau\gg\theta$.}
\label{fig4}
\end{figure}

\begin{figure}
\caption{The function $g(\tau(x))$ found by solving Eq.~(\protect\ref{gint})
numerically. Its dependence on the relaxation time
$\theta=\theta_0(\tau/\tau_0)^p$ is shown for various $\theta_0$ and $p$.}
\label{fig2}
\end{figure}

\begin{figure}
\caption{Entropy production in the relaxation time approximation
as function of the relaxation time $\theta=\theta_0(\tau/\tau_0)^p$
for various powers $p$. The entropy found by Baym \protect\cite{Baym}
(see text) is shown by dashed curve.}
\label{fig3}
\end{figure}

\end{document}